\documentclass[12pt]{article}
\usepackage[english]{babel}
\usepackage{graphicx,epsfig}
\makeindex \textwidth 170mm \textheight 220mm \topmargin -5mm
\newcommand{\be}{\begin{equation}}
\newcommand{\ee}{\end{equation}}
\newcommand{\bea}{\begin{eqnarray}}
\newcommand{\eea}{\end{eqnarray}}

\def\rfr#1{eq.(\ref{#1})}

\def\leti{Lense--Thirring}

\def\bb{\bibitem}
\def\eqi{\begin{equation}}
\def\eqf{\end{equation}}
\def\eqia{\begin{eqnarray}}
\def\eqfa{\end{eqnarray}}

\def\rp#1#2{{#1\over#2}}

\def\lb#1{\label{#1}}

\def\bm#1{{\mbox{\boldmath$#1$\unboldmath}}}

\begin{document}
\begin{center}
{\LARGE On the impossibility of measuring a galvano--gravitomagnetic
effect with current carrying semiconductors in a space--based experiment} \vspace{1.0cm}
\quad\\
{Lorenzo Iorio$^{\dag}$\\
\vspace{0.1cm}
\quad\\
{\dag}Dipartimento di Fisica dell' Universit{\`{a}} di Bari, via
Amendola 173, 70126, Bari, Italy\\ \vspace{0.1cm} }
\quad\\
 \vspace{1.0cm}

{\bf Abstract\\}
\end{center}

{\noindent \footnotesize In this paper we investigate the feasibility of a recently
proposed space--based experiment aimed to detect
the effect of the Earth's gravitomagnetic field in spaceborne
semiconductors carrying radial electric currents and following identical
circular equatorial orbits along opposite directions. Identical voltages of gravitomagnetic origin would be
induced across the orbiting semiconductors, while equal and opposite gravitoelctric larger voltages would
be generated. It turns out that the deviations
from the proposed idealized orbital configuration due to the unavoidable orbital injection errors would make impossible the
measurement of the gravitomagnetic voltage of interest.}
%\end{titlepage}

\newpage \pagestyle{myheadings} \setcounter{page}{1}
\vspace{0.2cm} \baselineskip 14pt

\setcounter{footnote}{0}
\setlength{\baselineskip}{1.5\baselineskip}
\renewcommand{\theequation}{\mbox{$\arabic{equation}$}}
\noindent \section{Introduction}
Up to now, the only proposed or performed attempts to measure some general relativistic
gravitomagnetic effects (Ciufolini and Wheeler 1995) in the space environment of Earth
by means of artificial satellites are
\begin{itemize}
  \item The well known Stanford GP--B mission (Everitt $et\ al$ 2001), which is aimed to detect the
  gravitomagnetic precession (Schiff 1960) of the spins of four superconductor gyroscopes and which is scheduled
  to fly in the second quarter of 2004
  \item The LAGEOS--LAGEOS II Lense--Thirring experiment (Ciufolini $et\ al$ 1998; Ciufolini 2002), which is currently performed by analysing the
orbital data of the existing LAGEOS and LAGEOS II geodetic laser--tracked satellites and
which measures the Lense--Thirring gravitomagnetic precessions (Lense and Thirring 1918) of the
longitude of the ascending node $\Omega$ and the argument of perigee $\omega$ of the orbits of the satellites
\item The proposed LAGEOS--LARES experiment, which is based on the
launch of a third LAGEOS--type satellite and which would measure the sum of the nodes of LAGEOS and LARES (Ciufolini 1986), or
a suitable combination of the nodes and the perigees of the three LAGEOS satellites (Iorio $et\ al$ 2002a). At present, funding
is the major obstacle to the implementation of such mission
\item The HYPER project, which is aimed to the detection of the
decoherence on microscopic probes induced by the Earth gravitomagnetic field by means of a space--based atom interferometer.
(${\rm http://www.esa.int/export/esaSC/SEM056WO4HD\_index\_0\_m.html}$) At present, such mission is under feasibility
assessment
\end{itemize}
Recently, in (Ahmedov 1999; 2003) another interesting experiment aimed to
the detection of a gravitomagnetic effect by means of a space based mission
has been proposed.
The analogy between the General Theory of Relativity, in its linearized weak--field and slow--motion approximation,
and electromagnetism suggests that there is a galvano-gravitomagnetic effect, which is the gravitational
analogue of the Hall effect. This new effect takes place when a current carrying conductor
is placed in a gravitomagnetic field and the conduction electrons moving inside the conductor are deflected
transversally with respect to the current flow.
In connection with this galvano-gravitomagnetic effect, in (Ahmedov 1999) the possibility of using
current carrying semiconductors for detecting the gravitomagnetic field of Earth by means of a pair of counter--orbiting satellites
has been outlined.

In this paper we wish to realistically analyze if such an
experiment could be effectively feasible. In Table \ref{para} we quote the physical quantities
of interest in the present case.
\begin{table}[ht!]
\caption{Physical quantities of interest in MKSA system.}
\label{para}
\begin{center}
\begin{tabular}{lll}
\noalign{\hrule height 1.5pt} Physical quantity & value & unit\\
\hline $G$ Newtonian gravitational constant & 6.67259$\times
10^{-11}$&
m$^3$ Kg$^{-1}$ s$^{-2}$\\
$J$ Earth's spin & 5.9$\times 10^{33}$ & Kg m$^2$
s$^{-1}$\\
$c$ speed of light in vacuum& 2.9978$\times 10^8$& m s$^{-1}$\\
$GM$ Earth $GM$ & 3.986$\times 10^{14}$& m$^3$ s$^{-2}$\\
$\delta (GM)$ error in Earth $GM$ (McCarthy 1996)& 8$\times 10^5$&  m$^3$
s$^{-2}$\\
$r_0$ geostationary satellite orbit radius& 4.2160$\times 10^7$&
m\\
$h$ difference of the satellites' orbit radius & 5$\times 10^3$ & m\\
$\delta r_0$ error in $r_0$& 10$^{-2}$& m\\
$\delta h$ error in $h$& 10$^{-2}$& m\\
$m_e$ electron rest mass& 9.11$\times 10^{-31}$& Kg\\
$e$ electron charge & $1.6\times 10^{-19}$ &C\\
$\nu_e$ semiconductor electronic concentration & $10^{18}$& m$^{-3}$\\
$i_r$ radial current carried by the spaceborne semiconductor &$10^3$& A\\
$d$ thickness of the spaceborne semiconductor & 10$^{-4}$& m\\
 \noalign{\hrule height 1.5pt}
\end{tabular}
\end{center}
\end{table}
%----------------------------------------------------------------
\section{The galvano--gravitomagnetic effect}
According to (Ahmedov 1999; 2003), restricting ourselves to circular equatorial
orbits, the gravitationally induced voltage across
a spaceborne semiconductor carrying a radial current $i_r$ and thickness $d$ is
given by the sum of a gravitoelectric contribution, which is sensitive to
the direction of motion along the satellite orbit, and a gravitomagnetic one, which, instead, is
insensitive to the direction of motion along the orbit\footnote{This situation is exactly reversed with respect to
the so called gravitomagnetic clock effect (Mashhoon $et\ al$ 1999; Iorio $et\ al$ 2002b).}
\eqi V=V_{\rm GE}+V_{\rm GM}=\pm \frac{n R_{gg}i_r}{d}+\frac{B_{g}
R_{gg}i_r}{2c d},\lb{voltage}\eqf where the sign + and - refer to the opposite
directions of motion of the satellites, $n$ is the Keplerian mean
motion \eqi n=(GM)^{\frac{1}{2}}r_0^{-\frac{3}{2}},\lb{kepmot}\eqf the
gravitomagnetic field is given, in general, by \eqi \bm
B_g=\frac{2GJ}{cr^3}[\bm {\hat{J}}-3(\bm {\hat J}\cdot\bm {\hat
r})\bm{\hat{r}}],\eqf and the galvano--gravitomagnetic constant is
\eqi R_{gg}=\frac{2m_e}{\nu_e e^2}.\eqf For an equatorial
geostationary orbit (see $r_0$ in Table \ref{para}) the
gravitomagnetic field, which is directed along the $z$ axis,
amounts to \eqi B_g=\frac{2GJ}{cr_0^3}=3.5\times 10^{-8}\ {\rm m}\
{\rm s}^{-2},\eqf while the galvano--gravitomagnetic constatnt,
for a typical semiconductor (see Table \ref{para}), amounts to
\eqi R_{gg}=7.117\times 10^{-11}\ {\rm Kg}\ {\rm m}^3\ {\rm
C}^{-2}.\eqf This implies that we have, with the values of Table
\ref{para}
\begin{eqnarray}\frac{R_{gg}i_r}{2cd}&\equiv&H=1.19\times 10^{-12}\ {\rm Kg}\ {\rm m}\ {\rm C^{-1}},\\
\frac{R_{gg}i_r}{d}&\equiv K&=7.117\times 10^{-4}\ {\rm Kg}\ {\rm
m^2}\ \ {\rm s^{-1}}\ {\rm C^{-1}},
\end{eqnarray} so that, for a geostationary
orbit, the gravitoelectric and gravitomagnetic voltages amount to \begin{eqnarray} V_{\rm GE}&\equiv& K n=
5.2\times 10^{-8}\ {\rm V},\\V_{\rm GM}&\equiv&H
B_g=4.16\times 10^{-20}\ {\rm V},\lb{voltaggio}\end{eqnarray} respectively. It
must be noted that an experimental sensitivity of
$10^{-12}$ V can be reached with great effort only by the today's SQUID technology;
a $10^{-15}$ V level, or even better, is still a challenge
(Vodel $et\ al$ 1995). This limits could be reached, e.g., by
increasing the intensity of the gravitomagnetic field $B_g$ felt
by the satellite by reducing its altitude. For a LAGEOS--type
orbit, i.e. a semimajor axis $a$ of 1.2270$\times 10^{7}$ m, the
gravitomagnetic field amounts to $1.4\times 10^{-4}$ m s$^{-2}$.
%---------------------------------------------------------------------
\section{The impact of the orbital injection errors}
Here we wish to consider in detail the effects of deviations of the spaceborne
semiconductors'orbits from the idealized case; we will put aside the problem of shielding the Earth magnetic field
which could induce much larger voltage due to the Hall effect.

Let us consider a couple of Earth artificial satellites following,
in principle, identical orbits around opposite directions and
denote them as $(+)$ and $(-)$: if this condition had been exactly
fulfilled, it turns out from \rfr{voltage} that in the sum of the
gravitationally induced voltages the gravitoelectric contributions
would cancel out while the gravitomagnetic ones would add up
\eqi\Sigma V\equiv V^{(+)}+V^{(-)}=\Sigma V_{\rm GE}+\Sigma V_{\rm
GM }=0+2V_{\rm GM}=2HB_{g}.\eqf Of course, rocketry is not an
exact science and the unavoidable orbital injection errors in the
orbit radius would make the paths followed by the
counter--orbiting satellites slightly different. This would induce
an uncancelled gravitoelectric component in the sum of the
voltages \eqi\Sigma V_{\rm GE}=K[n^{(+)}-n^{(-)}],\eqf so
that \eqi\Sigma V_{\rm GM}=\Sigma V-\Sigma V_{\rm GE}.\eqf Let us
pose $r_0^{(-)}=r_0^{(+)}+h$ with $h/r_0\ll 1$; then , from
\rfr{kepmot} it follows \eqi\Sigma V_{\rm GE}=K\rp{
3(GM)^{\frac{1}{2}}h r_0^{-\frac{5}{2}}}{2}.\eqf

Would the error in the gravitoelectric component $\delta(\Sigma
V_{\rm GE})$ be larger than the gravitomagnetic voltage of
interest $\Sigma V_{\rm GM}$? The error in $\Sigma V_{\rm GE}$ is
induced by the uncertainty in the Earth $GM$ and by the errors in
the Keplerian mean motions. In turn, they are induced by the
indirect effects due to $\delta r_0$ and $\delta h$ and by the
direct orbital perturbations on $n$ of gravitational and
non--gravitational origin. Let us consider the errors induced by
$\delta (GM), \ \delta r_0$ and $\delta h$. We have
\eqi\delta(\Sigma V_{\rm GE})\leq \left| K\rp{3 h
r_0^{-{\rp{5}{2}}}}{4(GM)^{\rp{1}{2}}}\right|\delta(GM)+\left| K
\rp{15(GM)^{\rp{1}{2}}h r_0^{-\rp{7}{2}}}{4}\right|\delta
r_0+\left|K\rp{3(GM)^{\rp{1}{2}}r_0^{-\rp{5}{2}}}{2}\right|\delta
h,\eqf from which it follows \eqi\rp{\delta (\Sigma V_{\rm GE})}{2
V_{\rm GM}}\leq \left|\rp{3c^2 h
r_0^{\rp{1}{2}}}{8G^{\rp{3}{2}}M^{\rp{1}{2}}J}\right|\delta(GM)+
\left|\rp{15 M^{\rp{1}{2}}c^2 h r_0^{-\rp{1}{2}} }{8 G
^{\rp{1}{2}}J}\right|\delta r_0+ \left|\rp{3 M^{\rp{1}{2}}c^2
r_0^{\rp{1}{2}} }{4 G ^{\rp{1}{2}}J}\right|\delta
h.\lb{errore}\eqf It is interesting to note that \rfr{errore}
depends on the Earth and satellite orbital parameters only.

According to the values of Table \ref{para}, the normalized error
of \rfr{errore} amounts to \eqi\rp{\delta (\Sigma V_{\rm GE})}{2
V_{\rm GM}}\leq 1\times 10^{-1}+6.6\times 10^{-2}+2.21\times
10^2.\eqf It can be noted that, while the errors due to $GM$ and
$r_0$ could be further reduced by increasing the accuracy of the
orbital injection process, i.e. with a smaller $h$, the major
limiting factor is given by the uncertainty in the knowledge of
the separation between the orbits of the two satellites. The value
$\delta h\sim 10^{-2}$ m is a conservative and realistic estimate;
for such a value it turns out that, unfortunately, there is no
hope of getting \eqi\left.\rp{\delta (\Sigma V_{\rm GE})}{2 V_{\rm
GM}}\right|_{\delta h}\leq 1\eqf by reducing the orbital
radius $r_0$. The situation does not change neither for $\delta
h\sim 10^{-3}$ m, which is a very stringent constraint, satisfied,
for example, in the GRACE mission.
%---------------------------------------------------
\section{Conclusions}
In this paper we have investigated the feasibility of a recently proposed space--based mission
aimed to the measurement of a gravitomagnetic effect on spaceborne semiconductors carrying radial currents
and following identical circular equatorial orbits along opposite directions.

It has been shown that the orbital injection errors in
the orbits of the proposed counter--rotating satellites would
induce a residual gravitoelectric voltage, which, instead, in an
idealized situation involving exactly equal orbits would be
cancelled. The associated uncertainty would be larger by one--two
orders of magnitude than the gravitomagnetic voltage of interest.
The main source of error would be the separation between the two
orbits.
%-----------------------------------------------------------------------------
\section*{Acknowledgements}
L Iorio warmly thanks B Ahmedov for his kind help in clarifying some important points.
Thanks also to L Guerriero for his support in Bari  and to the anonymous referee for his observations.
%-------------------------------------------------------------------------------------
%----------------------------------------------------------------
\end{document}